\RequirePackage{lineno}
\documentclass[a4paper,11pt]{article}
\pdfoutput=1 % if your are submitting a pdflatex (i.e. if you have images in pdf, png or jpg format)
\usepackage{jinstpub} % for details on the use of the package, please see the JINST-author-manual
\usepackage{multirow}
\usepackage{float}% for floating figures
\usepackage[font=small]{caption}
\usepackage[labelformat=empty, position=top]{subcaption}
\usepackage[export]{adjustbox}
\newcommand{\units}[1]{\hbox{$\,{\rm #1}$}}
\newcommand{\degrees}[0]{\hbox{${}^\circ$}}

\title{\boldmath Characterization of iLGADs using soft X-rays}
\author[a,1]{Antonio Liguori,\note{Corresponding author.}}
\author[a]{Rebecca Barten,}
\author[a]{Filippo Baruffaldi,}
\author[a]{Anna Bergamaschi,}
\author[b]{Giacomo Borghi,}
\author[b]{Maurizio Boscardin,}
\author[a]{Martin Br\"uckner,}
\author[a]{Tim Alexander Butcher,}
\author[a]{Maria Carulla,}
\author[b]{Matteo Centis Vignali,}
\author[a]{Roberto Dinapoli,}
\author[a]{Simon Ebner,}
\author[b]{Francesco Ficorella,}
\author[a]{Erik Fr\"ojdh,}
\author[a]{Dominic Greiffenberg,}
\author[b]{Omar Hammad Ali,}
\author[a]{Shqipe Hasanaj,}
\author[a]{Julian Heymes,}
\author[a]{Viktoria Hinger,}
\author[a]{Thomas King,}
\author[a]{Pawel Kozlowski,}
\author[a]{Carlos Lopez-Cuenca,}
\author[a]{Davide Mezza,}
\author[a]{Konstantinos Moustakas,}
\author[a]{Aldo Mozzanica,}
\author[b]{Giovanni Paternoster,}
\author[a]{Kirsty A. Paton,}
\author[b]{Sabina Ronchin,}
\author[a]{Christian Ruder,}
\author[a]{Bernd Schmitt,}
\author[a]{Dhanya Thattil,}
\author[a]{Xiangyu Xie,}
\author[a]{Jiaguo Zhang}
\affiliation[a]{Paul Scherrer Institut,\\Forschungsstrasse 111, 5232 Villigen, Switzerland}
\affiliation[b]{Fondazione Bruno Kessler,\\Via Sommarive 18, 38123 Trento, Italy}
\emailAdd{antonio.liguori@psi.ch}

\abstract{
Experiments at synchrotron radiation sources and X-ray Free-Electron Lasers in the soft X-ray energy range ($250\units{eV}$--$2 \units{keV}$) stand to benefit from the adaptation of the hybrid silicon detector technology for low energy photons. 
Inverse Low Gain Avalanche Diode (iLGAD) sensors provide an internal gain, enhancing the signal-to-noise ratio and allowing single photon detection below $1 \units{keV}$ using hybrid detectors.
In addition, an optimization of the entrance window of these sensors enhances their quantum efficiency (QE).

In this work, the QE and the gain of a batch of different iLGAD diodes with optimized entrance windows were characterized using soft X-rays at the Surface/Interface:Microscopy beamline of the Swiss Light Source synchrotron.  
Above $250\units{eV}$, the QE is larger than $55$\% for all sensor variations, while the charge collection efficiency is close to $100\%$. 
The average gain depends on the gain layer design of the iLGADs and increases with photon energy. 
A fitting procedure is introduced to extract the multiplication factor as a function of the absorption depth of X-ray photons inside the sensors. In particular, the multiplication factors for electron- and hole-triggered avalanches are estimated, corresponding to photon absorption beyond or before the gain layer, respectively.
}

\begin{document}
\maketitle
\flushbottom
%\linenumbers

\section{Introduction}
\label{sec:intro}

Hybrid detectors are widely used for the detection of hard and tender X-rays at synchrotron radiation sources and X-ray Free-Electron Lasers (XFELs) thanks to their versatility and robustness, combined with high dynamic range, fast frame rate, and the possibility of building large detector systems by tiling individual modules~\cite{detector_review}.  
Microscopy and spectroscopic techniques applied to the soft X-ray range ($250\units{eV}$ -- $2\units{keV}$) would also benefit from such characteristics to study material properties by exploiting the K absorption edges of light organic elements (e.g. C, N, O) or the L-edges of 3d transition metals (e.g. Cu, Fe, Ti).

\begin{figure}[h]

  \centering
  \begin{subfigure}[t]{0.015\textwidth}
    a)
  \end{subfigure}
  \begin{subfigure}[t]{0.47\textwidth}
    \includegraphics[width=\linewidth, valign=t]{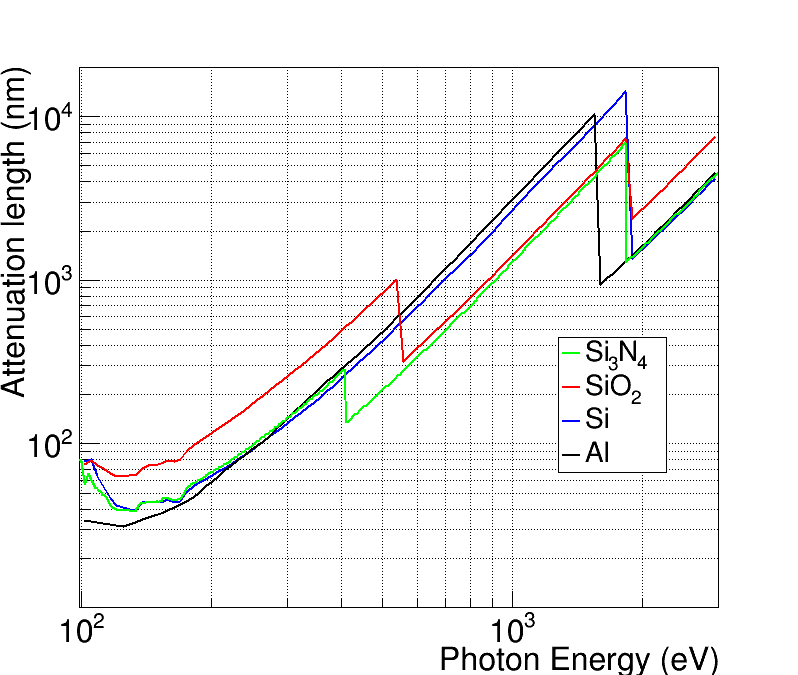}
  \end{subfigure}\hfill
  \begin{subfigure}[t]{0.015\textwidth}
    b)
  \end{subfigure}
  \begin{subfigure}[t]{0.45\textwidth}
    \includegraphics[width=\linewidth, valign=t]{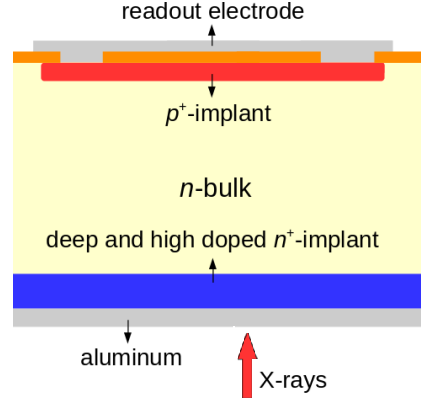}
  \end{subfigure}
  \caption{(a) Attenuation length as a function of the photon energy in  $\units{Al}$, $\units{Si_3N_4}$, $\units{SiO_2}$, and $\units{Si}$ between 100~eV and 3~keV~\cite{henke} and (b) cross section of a planar silicon sensor.}
  \label{fig:att_length}
\end{figure}

However, to make hybrid detectors suitable for applications in the soft X-ray energy range, the following challenges must be overcome:

\begin{itemize}
    \item The low quantum efficiency (QE) of planar silicon sensors below 2~keV, due to the sub-$\units{\mu m}$  attenuation length of soft X-rays in solids (figure~\ref{fig:att_length}(a)).
    Typical p-in-n silicon sensors for photon science (figure~\ref{fig:att_length}(b)) feature an X-ray entrance window composed of a $\sim1 - 2\units{\mu m}$ aluminum layer and $\sim1 - 2\units{\mu m}$ of highly doped n-type region.
    Soft X-rays can be easily absorbed in the insensitive aluminum layer. In addition, the charge carriers generated in the undepleted part of the n-type region may recombine before they generate a signal in the readout electrode. 
    \item The low signal-to-noise ratio (SNR) due to the small number of charge carriers created by soft X-rays. For example, $250\units{eV}$ photons produce $\sim$70~electron-hole pairs, which is comparable with the electronic noise of typical hybrid detectors. 
    Considering the noise of state-of-the-art hybrid detectors ($\sim30-100\;\text{e}^-$ equivalent noise charge RMS~\cite{viktoria, monch, medipix}), the minimal detectable energy of single X-ray photons with an SNR greater than 5 is $\sim600-1800\units{eV}$.
\end{itemize}

\begin{figure}[h]
\small
\centering
\includegraphics[width=140mm]{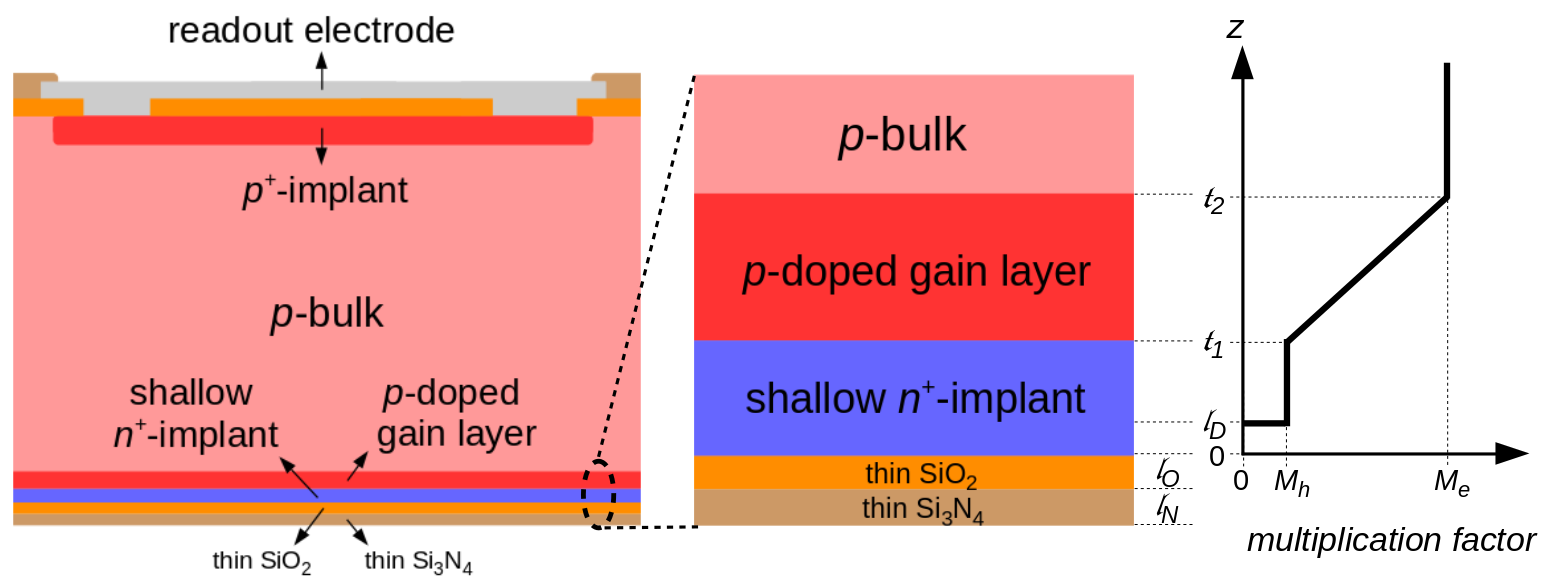}
\caption{Cross section of an iLGAD sensor with optimized EW.}
\label{fig:iLGAD}
\end{figure}

The Paul Scherrer Institut (PSI), in collaboration with Fondazione Bruno Kessler (FBK), is developing planar silicon sensors with optimized entrance windows (EW) for enhanced QE and, in parallel, inverse Low-Gain Avalanche Diode (iLGAD) sensors, to achieve single-photon detection in the soft X-rays range. The optimized EW has now been implemented on the iLGAD sensors. 

iLGADs~\cite{Carulla2016} are manufactured on p-type silicon wafers, with an $n^+$ contact and a $p$ gain layer on the backplane, where the EW is located, and readout channels formed segmenting a $p^+$ region on the front side (figure~\ref{fig:iLGAD}).
When an external reverse bias is applied, the electrons drift towards the backplane, while the holes are collected by the readout electrodes at the opposite side of the sensor.
The $p$-type gain layer is formed by boron implantation confined to within $1\units{\mu m}$ of the surface of the Si bulk. In this region, an electric field greater than 300~$\units{kV/cm}$ accelerates charge carriers so that they produce further carriers via impact ionization. The increase in the signal amplitude improves the SNR if the detector noise is not dominated by the shot noise in the sensor.
In iLGADs, the gain layer uniformly covers the entire sensor area. As a result, the fraction of the sensor area of a readout channel where the charge multiplication occurs is $100\%$, regardless of the segmentation of the readout electrodes.
This is an advantage compared to conventional LGADs, which have a no-gain region between segmented readout electrodes that is at least $30\units{\mu m}$ wide~\cite{ARCIDIACONO2020}, though this can be reduced in the range $2-5\units{\mu m}$ using trench-isolation~\cite{trench1, trench2}. The spatial uniformity of the multiplication is essential for X-ray position-sensitive detectors, in particular for experiments that require high spatial resolution by interpolation using charge sharing~\cite{sebastian}. The iLGAD sensors used in this work coupled to the M\"{O}NCH detector demonstrated single-photon detection down to a photon energy of $452\units{eV}$, when illuminated with fluorescence X-rays~\cite{jiaguo}. 

To optimize the EW, the metalization is replaced with a thin passivation, composed of $\units{Si_3N_4}$ and $\units{SiO_2}$ layers (figure~\ref{fig:iLGAD})~\cite{jiaguo}. In particular, the $\units{SiO_2}$ layer is used to reduce the recombination effects of charge carriers at the surface by compensating the silicon dangling bonds with the thermally grown oxide. The $n^+$ implant at the backplane is made as shallow as possible in order to reduce the width of the undepleted low-field region where charge carriers generated by X-rays can recombine. In this way, the charge carriers (holes in the iLGAD sensor) are able to diffuse out of the undepleted region and drift to the collecting electrode on the opposite side under the influence of a strong electric field. 
An investigation of the QE of planar silicon sensors with optimized EW using $405 \units{nm}$ UV light has been reported in~\cite{mar}. 

This paper presents measurements of the QE and the average gain of different variations of iLGADs combined with improved EW. The measurements span the photon energy range between $200 \units{eV}$ and $1\units{keV}$. Models describing the QE and average gain as functions of the energy are discussed and phenomenological models are also introduced for the charge multiplication factor as a function of the depth where the charge carriers are generated. The extracted depth-dependent multiplication factor is of 
fundamental importance to understand the spectral response of hybrid pixel detectors using iLGAD sensors to monoenergetic X-ray photons. It will be employed in device simulations to reproduce the measured spectral response of charge-integrating detectors, which will be discussed in a separate paper.
The investigation of the multiplication factor as a function of the photon absorption depth in the sensors under test and the study of their spectral response are relevant steps for the optimization of the design of iLGADs for soft X-ray detection using the hybrid detector technology. 

\section{Materials and methods}
\subsection{Sensors under test}
\label{sec:devices}

\begin{figure}
\small
\centering
\includegraphics[width=120mm]{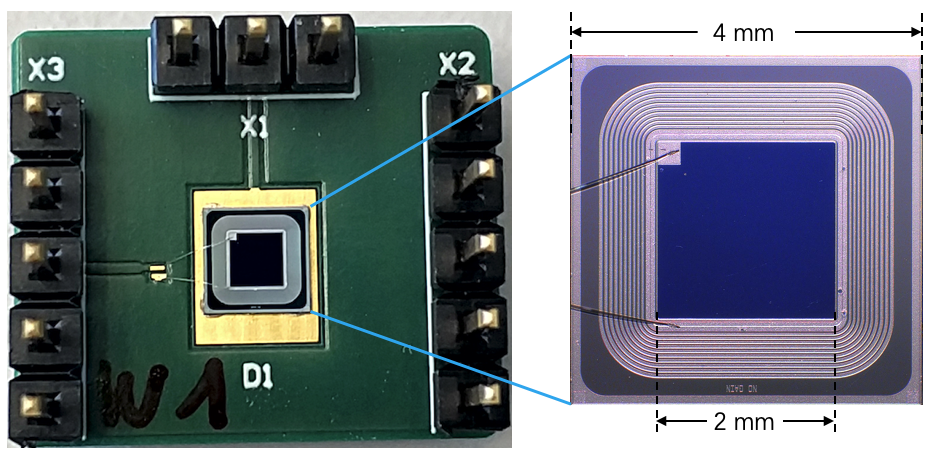}
\caption{Top view of the $n^{+}$-$p$-$p^{+}$ and iLGAD diodes under test. The active area is 4~mm$^{2}$ and is surrounded by one current-collection ring and ten floating guard rings.}
\label{fig:investigated_sensor}
\end{figure}

In order to characterize the quantum efficiency and gain of several sensor variations, pairs of X-ray sensing photodiodes, consisting of an iLGAD diode and an $n^{+}$-$p$-$p^{+}$ diode without gain layer manufactured by FBK, were tested. Both sensors were fabricated on the same silicon wafer and underwent the same production process, except for the implantation of the gain layer. The diodes are  $275\units{\mu m}$ thick and their active area is 4 mm$^{2}$,  surrounded by a current-collection ring (CCR) and ten floating guard rings to prevent breakdown at low bias voltages. Each sensor is glued onto a small PCB and wire-bonded to read out the photocurrent during the measurement (figure~\ref{fig:investigated_sensor}).

Among the process splits, two $n^+$-implant designs (shallow and ultra-shallow) at the EW and three gain layer profiles with different depths (namely standard, shallow, and ultra-shallow) and using different implantation doses have been investigated. Table~\ref{table:wafers} shows a summary of the $n^+$ and gain layer designs of the wafers under study. 

\begin{table}[t]
	\centering
	\begin{tabular}{|c|c|c|c|}
		\hline
        \hline
		Wafer identifier & $n^+$ implant & $p$ gain layer design & $p$ gain layer dose\\ 
        \hline
        \hline
		W15 & Standard & Standard & Low \\ \hline
		W17 & Shallow  & Standard & Medium\\ \hline  \hline
		W5  & Shallow  & Shallow   & Low \\ \hline
		W9  & Shallow  & Shallow   & High \\ \hline  \hline
		W19 & Ultra-shallow  & Ultra-shallow  & Medium  \\  \hline
		W13 & Ultra-shallow & Ultra-shallow & High \\
        \hline
        \hline
 \end{tabular}
	\caption{Characteristics of the different diodes under test as provided by the manufacturer.}
\label{table:wafers}
\end{table}

\subsection{Experimental setup}
\label{sec:expsetup}

\begin{figure}
\small
\centering
\includegraphics[width=150mm]{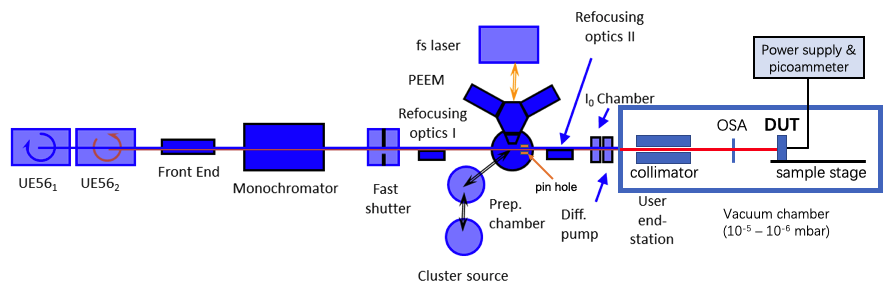}
\caption{The experimental setup at the SIM beamline of the Swiss Light Source. The picture is based on \cite{sim_layout}.}
\label{fig:setup}
\end{figure}

The measurements were carried out at the Surface/Interface:Microscopy (SIM) beamline of the Swiss Light Source (SLS) synchrotron. 
The beamline covers an energy range between $90 \units{eV}$ and $2\units{keV}$, with an X-ray energy resolution $E/\Delta E>5000$~\cite{simbeamline}. The experimental setup is shown in figure~\ref{fig:setup}.

The diodes under test were mounted on a motorized stage inside a vacuum chamber at the FLASH endstation of the beamline.
Three diodes fabricated by FBK were mounted at a time on the stage, along with a calibrated photodiode for which the quantum efficiency $QE_0(E)$ is known in the energy range of interest.
The temperature of the diodes was stabilized at $20 \degrees \units{C}$ with a liquid cooling system. The diodes were reverse biased at $300\units{V}$ simultaneously with a Keithley 6517B power supply and fully depleted.

The X-ray beam enters the vacuum chamber through a few$\units{mm}$ wide pinhole. It is collimated and then shaped by an Order Sorting Aperture (OSA), a $65 \units{\mu m}$ diameter pinhole positioned in front of the  diode under test. 
The stage could be moved to center the beam on the active area of each sensor. After proper alignment, the beam spot was fully contained in the active areas of the diodes. The OSA could be moved laterally to block the beam and measure the dark current. 

At each photon energy $E$,  the currents with X-ray illumination $I^\text{meas}$ and the dark currents $I^{\text{dark}}$ were measured for the three diodes under test and the calibrated photodiode. Four Keithley picoammeters were employed for the measurements. The photocurrent is calculated as $I^{\gamma}=I^{\text{meas}}-I^{\text{dark}}$ and is denoted as $I^\gamma_\text{pin}$, $I^\gamma_\text{iLGAD}$, and $I^\gamma_{0}$ 
for the $n^{+}$-$p$-$p^{+}$ diodes, the iLGAD diodes, and calibrated photodiode, respectively. 
The photocurrent of a diode depends on the photon energy $E$ and is proportional to the photon rate $\Phi$ of the X-ray beam. The fluctuations of $\Phi$ at a given photon energy during the experiments can be negletcted since the current of the electron beam in the synchrotron ring fluctuates by less than 0.25\%.

\subsection{Data analysis methods}

\subsubsection{Quantum efficiency}
\label{sec:quantumefficiency}
For the measurements of the QE of the iLGADs, we rely on the measurement of the $n^{+}$-$p$-$p^{+}$  diodes photocurrents, assuming that the QE is identical for devices fabricated on the same wafer with the same passivation and profile of the $n^{+}$ implant.
The QE of an X-ray sensing diode (without charge multiplication) is determined by the ratio between the photocurrent  and the current generated if all photons were absorbed inside the active region of the sensor and all carriers produced were collected:

\begin{equation}
QE=\frac{I^{\gamma}(E,\Phi)}{q_0\frac{\Phi E}{3.6\units{eV}}}\; ,
 \label{eq:qedef}
\end{equation}
\noindent where $q_0$ is the elementary charge, $E$ is the photon energy and $\Phi$ is the photon flux. 
Since this applies to both the $n^{+}$-$p$-$p^{+}$ diodes and the calibrated diode, the QE for each $n^{+}$-$p$-$p^{+}$  diode can be calculated by comparing $I^\gamma_\text{pin}$ and $I^{\gamma}_0$ at the same energy and flux and normalizing with the QE of the calibrated diode $QE_0$:

\begin{equation}
    QE_\text{pin}=\cfrac{I^{\gamma}_\text{pin}(E,\Phi)}{I^{\gamma}_0(E,\Phi)} \; QE_0 
 \label{eq:qecalc}
\end{equation}
For soft X-rays at normal incidence, the QE of the diode with $\units{SiO_2}$ and $\units{Si_3N_4}$ layers at the EW as a function of the photon energy can be expressed as: 

\begin{equation}
   QE(E) = \exp\left(-\frac{l_\text{N}}{\lambda_\text{N}(E)}\right)\exp\left(-\frac{l_\text{O}}{\lambda_\text{O}(E)}\right)\int_{0}^{L}CCE(z)\cdot \frac{1}{\lambda_{\text{Si}}(E)}\exp\left(-\frac{z}{\lambda_{\text{Si}}(E)}\right) dz  \; ,
     \label{eq:qemodel}
\end{equation}

\noindent where $l_\text{N}$ and $l_\text{O}$ are the thicknesses of the $\units{Si_3N_4}$ and $\units{SiO_2}$ layers; $\lambda_\text{N}$, $\lambda_\text{O}$, $\lambda_\text{Si}$ are the attenuation lengths of X-ray photons in $\units{Si_3N_4}$, $\units{SiO_2}$ and silicon and $L$ is the thickness of the silicon substrate. The first two factors on the right-hand side of equation~\ref{eq:qemodel} represent the transmission across the two dielectric layers, while the integral term corresponds to the average charge-collection efficiency ($CCE(z)$) at the given photon energy.
The $CCE(z)$ is defined as the fraction of the total photo-generated charge that is collected by the readout electrodes when the X-ray photon is absorbed at depth $z$ ~\cite{chargecollectionccd}. 
It can be assumed that the minimum value of the $CCE(z)$ occurs when a photon is absorbed at the surface ($z=0$), while it is unity for deep absorption, i.e. when a photon is absorbed within the depleted silicon region.

Regardless of the specific shape of $CCE(z)$, equation~\ref{eq:qemodel} allows a simple estimation of $l_\text{N}$ and $l_\text{O}$. An abrupt decrease in QE is expected at the K absorption edges of nitrogen and oxygen (i.e. at the energies $E_{\text{N}}=410\units{eV}$  and $E_{\text{O}}= 543\units{eV}$, respectively), due to the discontinuity of the attenuation lengths $\lambda_\text{N}(E)$ and $\lambda_\text{O}(E)$ (figure~\ref{fig:att_length}(a)). 
Taking into account the $\units{Si_3N_4}$ thickness $l_\text{N}$ first, it can be expressed, according to equation~\ref{eq:qemodel}, in terms of the ratio of the QE values at photon energies slightly above and below the nitrogen edge ($E_\text{N}^+$, $E_\text{N}^-$), :

\begin{equation}
  l_\text{N}  =  \ln \left({\frac{QE(E_\text{N}^+)}{QE(E_\text{N}^-)}} \right) \cdot \frac{\lambda_\text{N}(E_\text{N}^-)\lambda_\text{N}(E_\text{N}^+)}{\lambda_\text{N}(E_\text{N}^+)-\lambda_\text{N}(E_\text{N}^-)}
    \label{eq:thicknesseval1}
\end{equation}
\noindent The $\units{SiO_2}$ thickness $l_\text{O}$ can be expressed with an analogous formula, by considering the attenuation length $\lambda_\text{O}(E)$ instead of $\lambda_\text{N}(E)$ and by replacing the nitrogen edge $E_\text{N}$ with the one of oxygen $E_\text{O}$.

The simplest approach for describing of the $CCE(z)$ is the dead-layer model. It assumes that the charge carriers generated by an X-ray photon  absorbed within a thickness $l_\text{D}$ from the silicon surface, known as the dead layer, are completely lost due to recombination. Conversly, the charge is completely collected if the photon is absorbed beyond the dead layer. The corresponding expression of $CCE_{\text{dl}}(z)$ is:

\begin{equation}
  CCE_{\text{dl}}(z)=
\begin{cases}
0 \quad \text{if} \; 0 < z \leq l_\text{D}\\
1 \quad \text{if} \; z > l_\text{D}
\end{cases}
\label{eq:CCEdl}
\end{equation}

\subsubsection{Average gain and multiplication factor}
\label{sec:gain}

Assuming the same QE for both the iLGAD and the pin-diode coming from the same wafer, the average gain $g$ of an iLGAD diode can be defined as the ratio of the photocurrents from the iLGAD diode and the $n^{+}$-$p$-$p^{+}$ diode:  

\begin{equation}
    g(E)=\frac{I^{\gamma}_\text{iLGAD}(E,\Phi)}{I^{\gamma}_\text{pin}(E,\Phi)}. 
    \label{eq:gaindef}
\end{equation}

The multiplication factor $M$ is defined as the ratio between the number of electron-hole pairs that are collected at the boundaries of the depletion layer of the sensor and the initial number of electron-hole pairs introduced within it due to the absorption of an X-ray photon. 
Because of the difference in the impact ionization coefficient between electrons and holes in silicon, the multiplied charge depends on which charge carrier triggers the impact ionization and thus on the position where the X-ray photons are absorbed inside the sensor. 
In particular:

\begin{enumerate}
    \item when photons are absorbed in the $n^+$ implant close to the silicon surface, the multiplication is initiated by holes that drift through the gain layer and towards the $p^+$ electrode of the sensor; 
    \item if the absorption occurs beyond the gain layer the avalanche is instead triggered by electrons which  drift to the $n^+$-implant layer;
    \item if the carrier generation by X-ray photons occurs within the gain layer, the impact ionization is initiated by both primary electrons and holes.
\end{enumerate}

Because the impact ionization coefficient of electrons is larger than that of holes ($\alpha_{\text{e}} >\alpha_{\text{h}}$) \cite{impaction}, the multiplication factor $M_{\text{h}}$ for hole-initiated impact ionization (case 1) is lower than the multiplication factor $M_{\text{e}}$ for avalanches initiated by electrons (case 2). Therefore, the multiplication factor is a function $M(z)$  of the photon absorption depth $z$. As illustrated in figure~\ref{fig:iLGAD}, $M(z)=M_{\text{h}}$ for $z<t_1$, where $t_1$ is the onset of the gain layer, $M(z)=M_{\text{e}}$ for $z>t_2$, where $t_2$ is the end point of the gain layer, and $M(z)$ transitions from $M_{\text{h}}$ to $M_{\text{e}}$ ($M_{\text{e}}>M_{\text{h}}$) for $t_1<z<t_2$ (case 3). 

The average gain $g(E)$ and the multiplication factor $M(z)$ are related according to the equation:

 \begin{equation}
 g(E)=\frac{\int_{0}^{L}CCE(z)M(z)\frac{1}{\lambda_{\text{Si}}(E)}\exp\left(-\frac{z}{\lambda_{\text{Si}}(E)}\right) dz}{\int_{0}^{L}CCE(z)\frac{1}{\lambda_{\text{Si}}(E)}\exp\left(-\frac{z}{\lambda_{\text{Si}}(E)}\right) dz} \approx \int_{0}^{L}M(z) \frac{1}{\lambda_{\text{Si}}(E)}\exp\left(-\frac{z}{\lambda_{\text{Si}}(E)}\right) dz
 \label{eq:gmodel}
\end{equation}
In the last step of equation \ref{eq:gmodel}, it is assumed that the charge collection is complete (i.e. no charge loss inside the $n^{+}$ implant layer, $CCE\approx1$) and $L \gg \lambda_{\text{Si}}(E)$. If these conditions are fulfilled, the average gain corresponds to the mean value of the multiplication factor. A validation of the assumption $CCE\approx1$ will be given in section \ref{sec:qe_measurements}.
The average gain of an iLGAD diode is expected to increase with photon energy, because of the increasing probability of photon absorption beyond the gain layer.
A saturation of the average gain is expected at large $E$, when only a negligible fraction of photons is absorbed at $z<t_1$.

Due to the lack of suitable models that can accurately describe the ionization coefficients $\alpha_{\text{e}}(\mathcal{E})$ and $\alpha_{\text{h}}(\mathcal{E})$ in the range of electric field ($\mathcal{E}$) of LGAD sensors~\cite{moll}, it is difficult to express the multiplication factor analytically to calculate the average gain $g(E)$. Therefore, in this work, the following simple models for the multiplication factor as a function of the depth of absorption have been examined and inserted into equation~\ref{eq:gmodel}. The first model assumes a linear dependence of $M(z)$ on depth $z$:

\begin{equation}
  M_{\text{lin}}(z)=
\begin{cases}
M_{\text{h}} \quad \text{if} \; 0<z\leq t_1\\
\frac{M_{\text{e}}-M_{\text{h}}}{t_2-t_1}\cdot (z-t_1)+M_{\text{h}}\quad \text{if} \; t_1<z\leq t_2 \\
M_{\text{e}} \quad \text{if} \; t_2<z\leq L \quad.
\end{cases} 
\label{eq:linM}
\end{equation}
The second model considers an exponential transition of $M(z)$ from $t_1$ to $t_2$:

\begin{equation}
  M_{\text{exp}}(z)=
\begin{cases}
M_{\text{h}} \quad \text{if} \; 0<z\leq t_1\\
M_{\text{h}} \cdot \left(\frac{M_{\text{e}}}{M_{\text{h}}} \right) ^\frac{z-t_1}{t_2-t_1} \quad \text{if} \; t_1<z\leq t_2 \\
M_{\text{e}} \quad \text{if} \; t_2<z\leq L \quad.
\end{cases} 
\label{eq:exp1M}
\end{equation}

The multiplication factor as a function of depth, and hence the parameters $M_{\text{e}}$, $M_{\text{h}}$, $t_1$ and $t_2$, determine the spectral response of hybrid detectors coupled with iLGAD sensors to monochromatic X-rays.
The collected signal charge produced by an X-ray photon will depend on its absorption depth, where $M_{\text{h}}$ and $M_{\text{e}}$ respectively define the minimum and maximum charge that can be collected. For single photon detection at low energies only photons absorbed after the gain layer (electron-triggered multiplication, $M(z)=M_e$) will produce a SNR high enough to be detected. On the other hand, for multiple  photon detection, the broad distribution of the multiplication coefficients will result in a reduced accuracy in the conversion of the signal charge into number of photons, since only an average multiplication coefficient can be applied.
The relative probability of hole- and electron-triggered multiplication strongly depends on the bounds of the gain layer ($t_1$ and $t_2$), since in the soft X-ray energy range most of the X-ray photons are absorbed within the first $\mu m$ inside the active volume of the sensor below the EW.

\section{Results and discussion}
\label{sec:discussion}

\subsection{Quantum efficiency vs. photon energy}
\label{sec:qe_measurements}
\begin{figure}[h]
\small
\centering
\includegraphics[width=100mm]{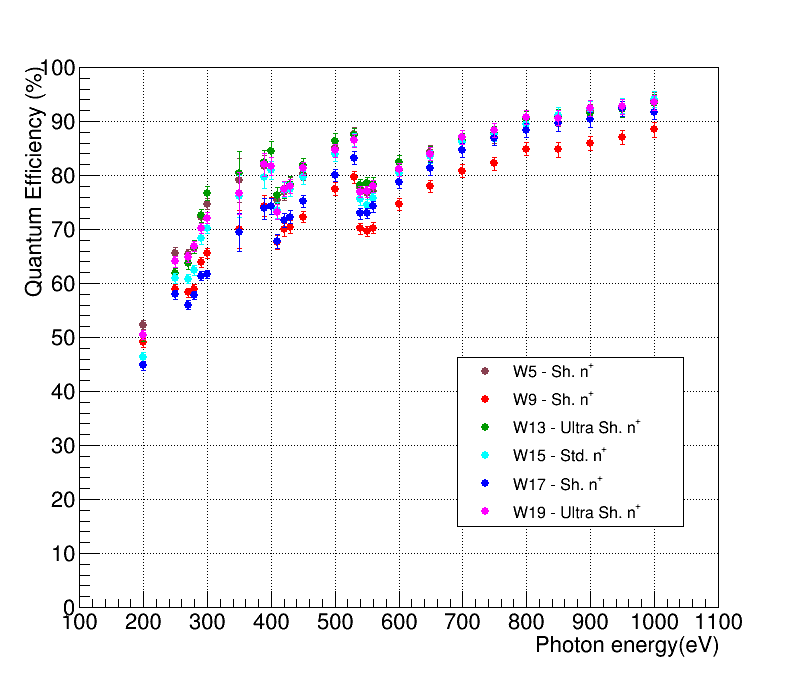}
\caption{QE as a function of the photon energy for the diodes under test listed in table~\ref{table:wafers}.}
\label{fig:QE}
\end{figure}

The QE as a function of photon energy obtained from the measurements using formula~\ref{eq:qecalc} is shown in figure~\ref{fig:QE} for all investigated diodes.
The QE values for photon energies above $250\units{eV}$ are larger than $55$\% for all variations.
This result represents a substantial improvement over conventional hard X-ray silicon sensors, which typically show a QE of the order of few \% at $250\units{eV}$ and less than $50\%$ at $800\units{eV}$ ~\cite{jiaguo}.

The increase in the attenuation length in $\units{Si_3N_4}$, $\units{SiO_2}$, and $\units{Si}$ with the photon energy, as shown in figure~\ref{fig:att_length}, results in a general trend of increasing QE at the higher photon energies. For X-ray photons with higher energies, absorption in the insensitive passivation layer and partial charge collection in the near-surface region as a result of recombination become less probable. In particular, at high energies the attenuation length of X-ray photons is significantly larger than the thicknesses of the dielectric layer and the $n^{+}$ implant for the different diodes under test; in this photon energy region the QE values for the different diodes are similar. 
The diode W9, which shows lower QE values compared to the others at high energies, is an exception. It is assumed that the different behavior is related to a faulty centering of the beam on the diode active area from the beginning: due to the shift of the motor position during the scan, part of the beam hits the border of the diode, which causes a smaller photocurrent and an underestimation of the QE. 

\begin{figure}[t]
 \centering
  \begin{subfigure}[t]{0.015\textwidth}
    a)
  \end{subfigure}
  \begin{subfigure}[t]{0.47\textwidth}
    \includegraphics[width=\linewidth, valign=t]{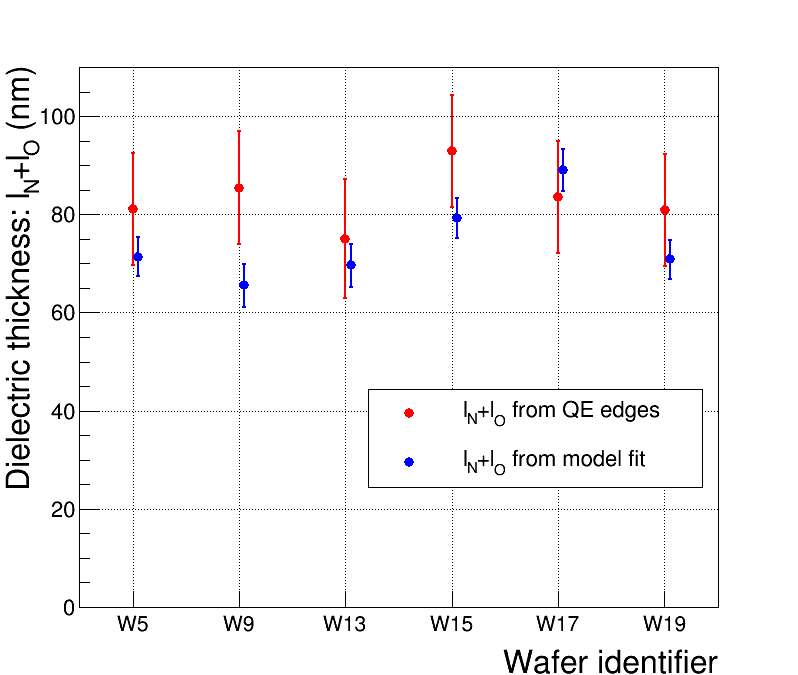}
  \end{subfigure}\hfill
  \begin{subfigure}[t]{0.015\textwidth}
    b)
  \end{subfigure}
  \begin{subfigure}[t]{0.47\textwidth}
    \includegraphics[width=\linewidth, valign=t]{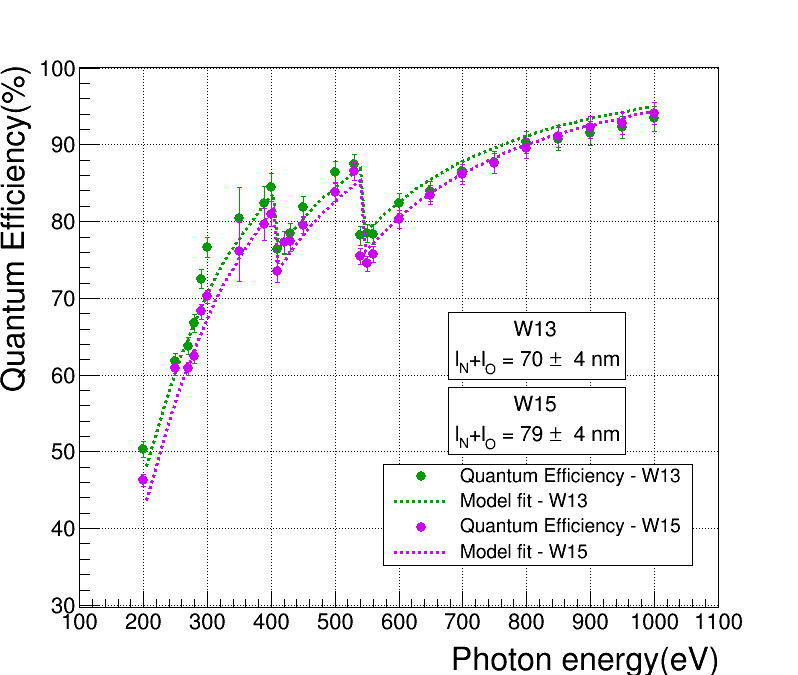}
  \end{subfigure}

\caption{(a) Total thickness of the dielectric layers, $l_\text{N}+l_\text{O}$, for different diodes. Red markers: evaluation from QE edges (equation~\ref{eq:thicknesseval1}); Blue markers: evaluation from the fit to the QE measurements using formula~\ref{eq:qemodel} with the dead layer model for $CCE(z)$. (b)  Fit of the QE measurements with formula~\ref{eq:qemodel} with the dead layer model for two representative wafers, W13 and W15.}
\label{fig:thickness_values}
\end{figure}

As expected, sudden drops in QE are present at the K absorption edges of nitrogen ($E_{\text{N}} \sim 410\units{eV}$) and oxygen ($E_{\text{O}} \sim 543\units{eV}$), due to the enhancement of the photon losses in the passivation.
The thicknesses of the $\units{Si_3N_4}$ and $\units{SiO_2}$ layers, $l_\text{N}$ and  $l_\text{O}$, are estimated using equation~\ref{eq:thicknesseval1}, where the QE measurements at the photon energies across the edges were used\footnote{For $l_\text{N}$ the QE points used are at $E_\text{N}^-=400\units{eV}$ and $E_\text{N}^+=410\units{eV}$, while for $l_\text{O}$ at $E_\text{O}^-=530\units{eV}$ and $E_\text{O}^+=550\units{eV}$.}. 
Since the individual thicknesses of the two layers cannot be disclosed, the total thickness $l_\text{N}+l_\text{O}$ is shown in figure~\ref{fig:thickness_values}(a) (red markers).

A fit of the QE measurements with expression~\ref{eq:qemodel} using the dead layer model $CCE_{\text{dl}}(z)$ was carried out with fitting parameters $l_N$, $l_O$, $l_D$.
For W9, the function in eq.~\ref{eq:qemodel} was multiplied by another parameter $A$ that represents the fraction of the beam impinging on the active area of the diode (best-fit value $A=0.904\pm0.009$).
Figure~\ref{fig:thickness_values}(b) shows QE functions fitted for two representative diodes (W13, W15). In all cases, the best-fit value of $l_{D}$ is $0 \units{nm}$, which indicates complete charge collection within the silicon bulk. The deterioration in QE is therefore dominated by photon losses in the inactive dielectric layers of the EW.

In figure~\ref{fig:thickness_values}(a), the sum of $l_\text{N}+l_\text{O}$ obtained from the fit to the data from different diodes is plotted in blue.
The fit of the data with the function in equation~\ref{eq:qemodel} yields $5-10\units{nm}$ lower values of $l_\text{N}+l_\text{O}$ compared to the values obtained from the measurement using the QE drops at the K absorption edges. Increasing the fit parameters $l_\text{N}$ and $l_\text{O}$ in the QE model to match the amplitudes of the measured QE edges results in an overall reduction of the QE and, thus, increases the residuals. The thicknesses obtained for the dielectric layers using the two different methods are consistent within the statistical error. 

\subsection{Average gain vs. photon energy}
\label{sec:gain_measurements}

\begin{figure}
\small
\centering
\includegraphics[width=100mm]{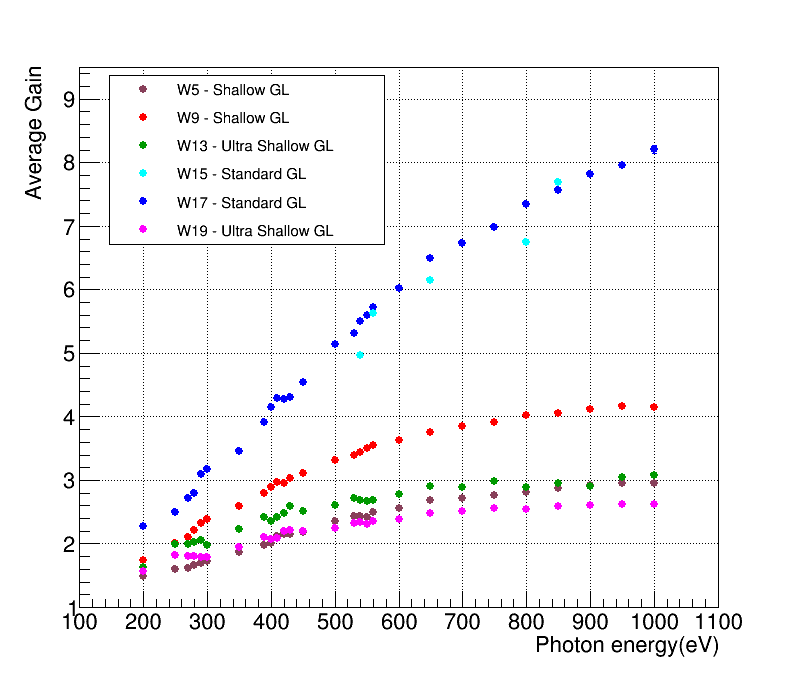}
\caption{Average gain as a function of the photon energy for the iLGAD diodes listed in table \ref{table:wafers}. Due to the large fluctuation of current measured for W15 iLGAD, only the results with a relative error below 3.5\% are shown for this wafer. }
\label{fig:gain}
\end{figure}

Figure~\ref{fig:gain} displays the average gain of the iLGAD diodes (equation \ref{eq:gaindef}).
It can be seen that the average gain increases with photon energy, due to the increase of the probability of absorption beyond the gain layer. As discussed in section~\ref{sec:gain}, this corresponds to electron-initiated avalanches, which are characterized by a higher multiplication factor compared to the hole-initiated ones ($M_{\text{e}}>M_{\text{h}}$). 

Because of the thicker gain layer, diodes featuring the standard gain layer design (W15, W17) exhibit higher average gain than diodes with shallow and ultra-shallow designs.
The saturation of the average gain at high $E$ is obtained when only a small fraction of the impinging photons is absorbed before or within the gain layer. 
The saturation is reached for ultra-shallow gain layer designs (W13, W19), while for the standard gain layer (W15, W17) $g(E)$ continues to increase beyond the upper bound of the photon energy in the measurement.

Different average gain values can be observed both between the two iLGADs with shallow gain layer designs (W5 and W9) and between the two with ultra-shallow designs (W19 and W13). These discrepancies can be attributed to lower doses of the $p$-type gain layer in W5 and W19 (table \ref{table:wafers}). This results in a reduced electric field intensity in the multiplication region, leading to lower impact ionization coefficients of both electrons and holes.

According to the results of section~\ref{sec:qe_measurements}, the average gain of the iLGADs as a function of energy can be fitted using equation~\ref{eq:gmodel} and assuming $CCE(z)\approx 1$. From the fit, the multiplication factors of electron-initiated avalanches and hole-initiated avalanches ($M_{\text{e}}$ and $M_{\text{h}}$) can be extracted, as described in section \ref{sec:mult_meas}.

\subsection{Multiplication factor vs. absorption depth}
\label{sec:mult_meas}
The models $M_{\text{lin}}(z)$ and $M_{\text{exp}}(z)$ were used in equation~\ref{eq:gmodel} for the average gain fit, with free parameters $M_{\text{e}}$, $M_{\text{h}}$, $t_2$. The parameter $t_1$ was fixed at the onset of the depletion layer under full depletion condition. Its value is determined by solving a one-dimensional Poisson equation that takes into account the doping profiles of the $n^{+}$ implant and the $p$-type gain layer and the doping concentration of the substrate. The boundary conditions for the solution are an electric potential of 0 V at the $p^{+}$ readout electrode and 300 V at the $n^{+}$ contact. The calculated value of $t_1$ approximates the onset of the region where the electric field is of the order of $\sim300\units{kV/cm}$ and multiplication occurs.

\begin{figure}[t!]
 \centering
  \begin{subfigure}[t]{0.015\textwidth}
    a)
  \end{subfigure}
  \begin{subfigure}[t]{0.47\textwidth}
    \includegraphics[width=\linewidth, valign=t]{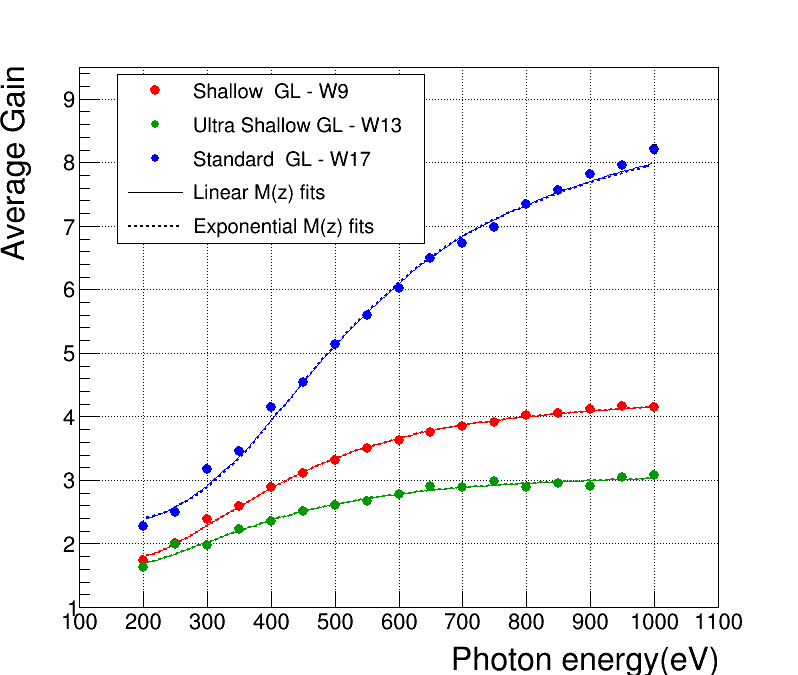}
  \end{subfigure}\hfill
  \begin{subfigure}[t]{0.015\textwidth}
    b)
  \end{subfigure}
  \begin{subfigure}[t]{0.47\textwidth}
    \includegraphics[width=\linewidth, valign=t]{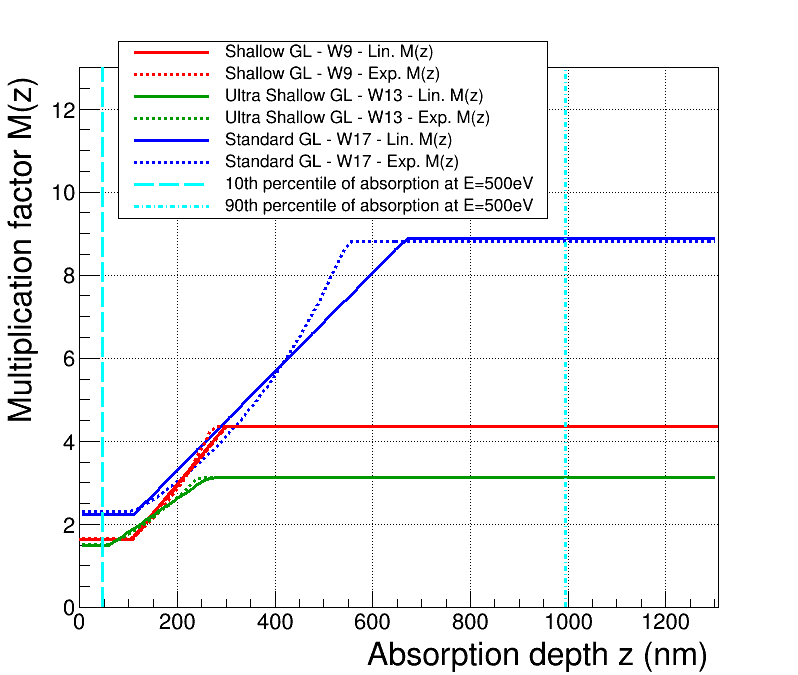}
  \end{subfigure}
    \caption{(a) Fits of the average gain $g(E)$ for the iLGADs W9, W13 and W17. The fits were carried out with equation~\ref{eq:gmodel}, using the linear model (equation~\ref{eq:linM}) and the exponential model (equation~\ref{eq:exp1M}) for $M(z)$. The best-fit functions are similar for both models.
    (b) The models $M_{\text{lin}}(z)$ (solid lines) and $M_{\text{exp}}(z)$ (dotted lines) with best-fit parameters $M_{\text{e}}$, $M_{\text{h}}$, $t_2$ for the iLGADs W9 (red), W13 (green), W17 (blue); the 10th and 90th percentiles of the absorption depth for $500\units{eV}$ photons (cyan).}
	\label{fig:gainfit}
\end{figure}

In figure~\ref{fig:gainfit}(a) the fits to the average gain data are shown, while the best-fit parameters are reported in table~\ref{table:gainparamtab}. It can be seen that the best-fit functions using the linear and exponential models are almost overlapping, indicating that the measurements of the average gain do not favor one model over the other. 
Figure \ref{fig:gainfit}(b) shows the dependence of the multiplication factor on the depth of absorption for the best-fit linear models and exponential models.
The extracted $M_{\text{e}}$ and $M_{\text{h}}$ for both models are consistent. 
The exponential increase of the multiplication factor in $M_{\text{exp}}(z)$ is compensated by a smaller $t_2$ compared to $M_{\text{lin}}(z)$ and yields a similar $M_{\text{e}}$.
In order to study the sensitivity of the best-fit parameters to the fixed value of $t_1$, the fits were repeated by increasing or decreasing $t_1$ by $10\%$ (6.1--11~nm, depending on the iLGAD design). 
Considering the linear model, the best-fit values of $M_\text{h}$ change by $0.80-1.2\%$ (0.012--0.031, depending on the iLGAD), $M_\text{e}$ changes by $0.016-0.14\%$ (0.005--0.012), and $t_2$ varies by $1.8-3.6\%$ (4.7--14~nm).  Similar results are obtained with the exponential model of $M(z)$.

According to the fitted parameters (table~\ref{table:gainparamtab}), the standard gain layer design (W17) provides larger multiplication factors $M_{\text{e}}$ and $M_{\text{h}}$ than the shallow (W9) and ultra-shallow (W13) designs. On the other hand, these factors do not scale with the thickness of the gain layer, $t_2-t_1$, since they depend on the specific electric field profile within the multiplication region as well.
The ratio $M_{\text{h}}/M_{\text{e}}$ is higher in the ultra-shallow (W13) and shallow (W9) designs  than the standard design (W17), due to the different dependence of the impact ionization coefficients of electrons ($\alpha_{\text{e}}$) and holes ($\alpha_{\text{h}}$) on the electric field \cite{moll}. The electric field in the gain layer for W13 and W9 is higher than that for W17, resulting in a higher ratio $\alpha_{\text{h}} / \alpha_{\text{e}}$ and thus higher $M_{\text{h}}/M_{\text{e}}$. 

\begin{table}[t]
\begin{tabular}{|c|c|c|c|c|c|c|c|}
\hline
   \textbf{Device} & $p$-implant & $M(z)$ & $M_{\text{e}}$        & $M_{\text{h}}$         & $t_1 (\units{nm})$ & $t_2 (\units{nm})$ & $M_{\text{h}}/M_{\text{e}}$ \\ \hline \hline

\multirow{ 2}{*}{\textbf{W17}} & Standard & Lin  & $8.88\pm0.03$ & $2.25\pm0.02$ & \multirow{ 2}{*}{$112$} & $670\pm12$ & $0.253 \pm 0.003$\\ \cline{3-5}\cline{7-8}%\hline
 & Medium & Exp  & $8.81\pm0.03$ & $2.31\pm0.02$ & & $548\pm10$ & $0.262 \pm 0.003$\\ \hline \hline

\multirow{ 2}{*}{\textbf{W9}} & Shallow & Lin & $4.355\pm0.006$ & $1.64\pm0.01$ & \multirow{ 2}{*}{$107$} & $299\pm5$ & $0.377 \pm 0.003$\\ \cline{3-5}\cline{7-8}%\hline
 & High & Exp  & $4.352\pm0.006$ & $1.65\pm0.01$ & & $271\pm5$ & $0.380 \pm 0.003$\\ \hline \hline

\multirow{ 2}{*}{\textbf{W13}} & UltraShallow & Lin  & $3.12\pm0.01$ & $1.50\pm0.04$ & \multirow{ 2}{*}{$61$} & $263\pm18$ & $0.48 \pm 0.01$\\ \cline{3-5}\cline{7-8}%\hline
& High & Exp  & $3.12\pm0.01$ & $1.51\pm0.04$ &  & $242\pm17$ & $0.49 \pm 0.01$\\ \hline \hline
   
\end{tabular}

\caption{Parameters of the linear and exponential models $M_{\text{lin}}$ and $M_{\text{exp}}$ from the fits of the average gain for the iLGAD diodes W9, W13 and W17. $t_1$ is a fixed parameter in the fits. The iLGADs were biased at $300\units{V}$ and their temperature was stabilized at $+20^\circ\units{C}$.}
\label{table:gainparamtab}
\end{table}  

\section{Summary and outlook}
\label{sec:summary}

The adaptation of the hybrid detector technology for soft X-ray detection involves the development of iLGAD sensors with an optimized EW to improve both the quantum efficiency (QE) and the signal-to-noise ratio (SNR). 
Variations of diodes of this type with and without gain, fabricated by FBK, have been characterized using soft X-rays in the photon energy range between $200\units{eV}$ and $1\units{keV}$ at the SIM beamline of the SLS synchrotron.

The QE is in the range $55-67$\% at $250\units{eV}$ for all process variations, a significant improvement compared to conventional planar sensors. 
Additionally, the QE increases with photon energy due to the increase of the attenuation length, with discontinuities at the K absorption edges of nitrogen and oxygen, because of the use of  $\units{SiO_2}$ and $\units{Si_3 N_4}$ dielectric layers in the EW.
The dependence of the QE on photon energy can be described with a model that considers photon absorption in the dielectric layers and carrier recombination in a dead layer of silicon close to the silicon surface.  
With this model, we showed that the dominant cause of QE degradation is photon loss in the dielectric layers (total thickness $\approx 70-90 \units{nm}$ ), while the charge collection is almost $100\%$ inside the silicon sensor ($CCE\approx1$). 
This study suggests that the thicknesses of the dielectric layers need to be reduced, without sacrificing their uniformity over a large area, for further improvement of the QE.

For all iLGAD diodes studied, the average gain $g(E)$ increases with photon energy.
This behavior is attributed to the increased likelihood of electron-initiated multiplication (corresponding to absorption beyond the gain layer) at higher photon energies.
The standard gain layer design yields a higher average gain with respect to the shallow and ultra-shallow ones. In addition, iLGADs with a higher dose in the p-type gain layer (with shallow or ultra-shallow designs) exhibit a higher average gain.

Two empirical models were introduced for $g(E)$, which involve a linear and an exponential transition of the multiplication factor $M(z)$ as a function of the absorption depth $z$ from the value $M_{\text{h}}$ (corresponding to $z<t_1$ i.e. absorption before the gain layer and hole-initiated avalanches) to $M_{\text{e}}$ (corresponding to $z>t_2$, i.e. absorption beyond the gain layer and electron-initiated avalanches). 
Both models describe the measured $g(E)$ well and the fitted values of $M_{\text{h}}$ and $M_{\text{e}}$ turn out to be independent of the model chosen.
The standard gain layer design produces larger $M_{\text{h}}$ and $M_{\text{e}}$, potentially leading to an higher SNR for single photon detection using hybrid pixel detectors in comparison to the shallower designs. 
On the other hand, the shallow and ultra-shallow designs take advantage from a higher probability of electron-triggered multiplication  (lower $t_2$ value) at a given photon energy.
This property may provide a higher photon detection efficiency at low energy, if the SNR associated with electron-triggered multiplication is sufficient to achieve single photon detection.
In addition, the ratio $M_{\text{h}}/M_{\text{e}}$ is higher for the shallow and ultra-shallow gain layer designs, because of the higher the electric field intensity in the multiplication region, compared to the standard one. A higher ratio $M_{\text{h}}/M_{\text{e}}$ could be an advantage for the operation of both single-photon counting and charge-integrating hybrid detectors, since it results in more similar amplitudes of the output signals due to hole- and electron-triggered multiplication of the charge produced by an X-ray photon.

The fitting procedure presented in this work provides a way to estimate both $M_{\text{e}}$ and $M_{\text{h}}$ from the measurements of the average gain of the iLGADs. It is valuable for future investigation of their dependence on electric field and temperature.
Further studies of $M(z)$ in iLGAD sensors will entail measuring single photon spectra of hybrid pixel charge-integrating detectors (e.g. M\"{O}NCH~\cite{monch}  and JUNGFRAU~\cite{viktoria}) coupled with the same iLGADs used in this work.
The extracted parameters of the linear and exponential models for $M(z)$ will serve as a basis for simulations of these spectra. By comparing the simulation results with experimental data, further insight will be also obtained into the transition of $M(z)$ between $M_{h}$ and $M_{e}$.
\vfill

\acknowledgments
Two of the authors, V. Hinger and K. A. Paton, have received funding from MSCA PSI-FELLOW-III-3i (EU grant agreement No. 884104). T. A. Butcher acknowledges funding from the Swiss Nanoscience Institute (SNI). Measurements were performed at the Surface/Interface: Microscopy (SIM) beamline of the Swiss Light Source, Paul Scherrer Institut, Villigen PSI, Switzerland. The authors thank Armin Kleibert of the SIM beamline for the allocation and coordination of the beam time, J\"org Raabe and Carlos Antonio Fernandes Vaz for the support during the beam time, Thomas Huthwelker and Camelia Nicoleta Borca of the PHOENIX beamline for providing the calibrated photodiode and its calibration data, used extensively in this study. A. Liguori would like to thank Prof. Francesco Loparco from the University of Bari Aldo Moro and the Detector group of the Photon Science Division of PSI for offering the opportunity to work on this research topic for his master's thesis.

\end{document}